\documentclass{sig-alternate-10pt}
\usepackage[margin=1in]{geometry}
\usepackage{tikz}

\usepackage{multicol}
\usepackage{color}
\usepackage{xspace}
\usepackage{enumerate}
\usepackage{color}
\usepackage{url}

\usepackage{multirow}
\usepackage{graphicx}

\usepackage{graphicx}
\graphicspath{{./figure/}}
\usepackage{subfig}

\usepackage{hyphenat}
\usepackage[utf8]{inputenc}

\usepackage{cleveref}
\crefname{section}{§}{§§}

\newcommand{\compact}{\vspace{-4pt}}

\makeatletter
\newcommand\subparagraph{%
  \@startsection{subparagraph}{5}
  {\parindent}
  {3.25ex \@plus 1ex \@minus .2ex}
  {-1em}
  {\normalfont\normalsize\bfseries}}
\makeatother
\usepackage{titlesec}
\titlespacing{\section}{4pt}{4pt}{4pt}
\titlespacing{\subsection}{4pt}{3pt}{3pt}
\titlespacing{\subsubsection}{4pt}{2pt}{2pt}

\begin{document}
\pagestyle{empty}

\title{Hardware-Conscious Stream Processing: A Survey}
 
\author{
{Shuhao Zhang$^{1}$, Feng Zhang$^2$, Yingjun Wu$^3$, Bingsheng He$^1$, Paul Johns$^1$}\\
$^{1}$National University of Singapore,  $^2$Renmin University of China,\\
$^{3}$Amazon Web Services
} 


\maketitle


\begin{abstract}
Data stream processing systems (DSPSs) enable users to express and run stream applications to continuously process data streams. 
To achieve real-time data analytics, recent researches keep focusing on optimizing the system latency and throughput.
Witnessing the recent great achievements in the computer architecture community, researchers and practitioners have investigated the potential of adoption hardware-conscious stream processing by better utilizing modern hardware capacity in DSPSs. 
In this paper, we conduct a systematic survey of recent work in the field,
particularly along with the following three directions: 1) computation optimization, 2) stream I/O optimization, and 3) query deployment. 
Finally, we advise on potential future research directions.
\end{abstract}

\compact
\section{Introduction}
\label{sec:introduction}
A large volume of data is generated in real time or near real time and has grown explosively in the past few years.
For example, IoT (Internet-of-Things) organizes billions of devices around the world that are connected to the Internet. 
IHS Markit forecasts~\cite{iot_report} that 125 billion such devices will be in service by 2030, up from 27 billion in 2018.
With the proliferation of such high-speed data sources,
numerous data-intensive applications are deployed in real-world use cases exhibiting latency and throughput requirements, that can not be satisfied by traditional batch processing models. 
Despite the massive effort devoted to big data research, many challenges remain.

A data stream processing system (DSPS) is a software system which allows users to efficiently run stream applications that continuously analyze data in real time.
For example, modern DSPSs~\cite{flink, Storm} can achieve very low processing latency in the order of milliseconds.
Many research efforts are devoted to improving the performance of DSPSs from the research community~\cite{SABER, CellJoin, Analyzing, ChronoStream} and leading enterprises such as SAP~\cite{motto}, IBM~\cite{system-s}, Google~\cite{millwheel} and Microsoft~\cite{trill}.
Despite the success of the last several decades, more radical performance demand, complex analysis, as well as intensive state access in emerging stream applications~\cite{ingestion17,8509389} pose great challenges to existing DSPSs.
Meanwhile, significant achievements have been made in the computer architecture community, which has recently led to various investigations of the potential of \emph{hardware-conscious DSPSs}, which aim to exploit the potential of accelerating stream processing on modern hardware~\cite{briskstream, Analyzing}.

Fully utilizing hardware capacity is notoriously challenging.
A large number of studies have been proposed in recent years~\cite{trill, TerseCades, StreamBox, stream_hbm, SABER, Analyzing, profile, briskstream}.
This paper hence aims at presenting a systematic review of prior efforts on hardware-conscious stream processing.
Particularly, the survey is organized along with the following three directions: 1) computation optimization, 2) stream I/O optimization, and 3) query deployment.
We aim to show what has been achieved and reveal what has been largely overlooked.
We hope that this survey will shed light on the hardware-conscious design of future DSPSs.

\compact
\section{Background}
\label{sec:bg}
In this section, we introduce the common APIs and runtime architectures of modern DSPSs. 

\subsection{Common APIs}
\label{subsec:api} 
\begin{figure}[t]
\centering
        \includegraphics[width=0.4\textwidth]{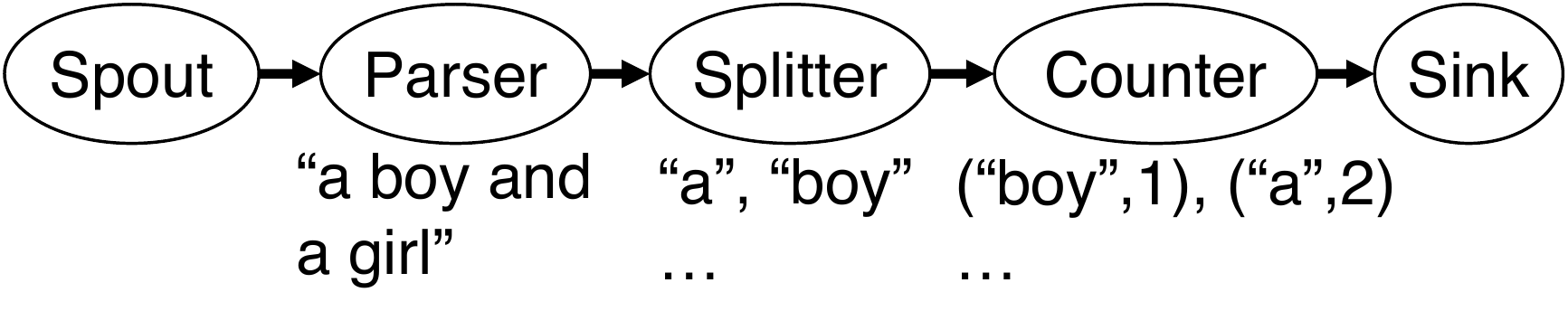}    
    \caption{A stream processing example of word count.}                   
    \label{fig:example}       
\end{figure}

A DSPS needs to provide a set of APIs for users to express their stream applications. 
Most modern DSPSs such as Storm~\cite{Storm} and Flink~\cite{flink} express a streaming application as a directed acyclic graph (DAG), where nodes in the graph represent operators, and edges represent the data dependency between operators. 
Figure~\ref{fig:example} (a) illustrates the \emph{word count} (WC) as an example application containing five operators. 
A detailed description of a few more stream applications can be found in~\cite{profile}.


Some earlier DSPSs (e.g., Storm~\cite{Storm}) require users to implement each operator manually. 
Recent efforts from Saber~\cite{SABER}, Flink~\cite{flink}, Spark-Streaming~\cite{SparkStreaming}, and Trident~\cite{trident} aim to provide declarative APIs (e.g., SQL) with rich built-in operations such as aggregation and join.
Subsequently, many efforts have been devoted to improving the execution efficiency of the operations, especially by utilizing modern hardware (Section~\ref{sec:operator}).

\subsection{Common Runtime Architectures}
\label{subsec:archi}
Modern stream processing systems can be
generally categorized based on their processing models including the
Continuous Operator (CO) model and the Bulk Synchronous Parallel (BSP) model~\cite{Drizzle}. 

\emph{Continuous Operator Model:}
Under the CO model, the execution runtime treats each operator (a vertex of a DAG) as a single execution unit (e.g., a Java thread), and multiple operators communicate through message passing (an edge in a DAG). 
For scalability, each operator can be executed independently in multiple threads, where each thread handles a substream of input events with stream partitioning~\cite{partition}. 
This execution model allows users to control the parallelism of each operator in a fine-grained manner~\cite{profile}.
This kind of design was  adopted by many DSPSs such as Storm~\cite{Storm}, Heron~\cite{heron}, Seep~\cite{CastroFernandez:2013:ISO:2463676.2465282}, 
and Flink~\cite{flink} due to its advantage of low processing latency.
Other recent hardware-conscious DSPSs adopt the CO model including Trill~\cite{trill}, BriskStream~\cite{briskstream}, and TerseCades~\cite{TerseCades}. 

\emph{Bulk-Synchronous Parallel Model:}
Under the BSP model, input stream is explicitly grouped into micro batches by a central coordinator and then distributed to multiple workers (e.g., a thread/machine). 
Subsequently, each data item in a micro batch is independently processed by going through the entire DAG (ideally by the same thread without any cross-operator communication).
However, the DAG may contain synchronization barrier, where threads have to exchange their intermediate results (i.e., data shuffling).
Taking WC as an example, the \texttt{Splitter} needs to ensure that the same word is always passed to the same thread of the \texttt{Counter}. 
Hence, a data shuffling operation is required before the \texttt{Counter}.
As a result, such synchronization barriers break the DAG into multiple stages under the BSP model, and the communication between stages is managed by the central coordinator.
This kind of design was  adopted by Spark-streaming~\cite{SparkStreaming}, Drizzle~\cite{Drizzle}, and FlumeJava~\cite{FlumeJava}. 
Other recent hardware-conscious DSPSs adopt the BSP model including Saber~\cite{SABER} and StreamBox~\cite{StreamBox}.

Although there have been prior efforts to compare different models~\cite{flow}, it is still inconclusive that which model is more suitable for utilizing modern hardware -- each model comes with its own advantages and disadvantages.
For example, the BSP model naturally minimizes communication among operators inside the same DAG, but its single centralized scheduler has been identified with scalability limitation~\cite{Drizzle}. 
Moreover, its unavoidable data shuffling also brings significant communication overhead, as observed in recent research~\cite{briskstream}.
In contrast, CO model allows fine-grained optimization (i.e., each operator can be configured with different parallelisms and placements) but potentially incurs higher communication costs among operators. 
Moreover, the limitations of both models can potentially be addressed with more advanced techniques.
For example, cross operator communication overhead (under both CO and BSP models) can be overcome by exploiting tuple batching~\cite{trill,profile}, high bandwidth memory~\cite{stream_hbm,stream_hbm2}, 
data compression~\cite{TerseCades}, InfiniBand~\cite{Heron_Infiniband} (Section~\ref{sec:flow}), and architecture-aware query deployment~\cite{briskstream,Analyzing} (Section~\ref{sec:query}).

\compact
\begin{table*}[t]
\centering
\scriptsize
\caption{Summary of the surveyed works}
\label{tbl:summary}
\resizebox{\textwidth}{!}{%
\begin{tabular}{|p{0.11\linewidth}|p{0.25\linewidth}|p{0.45\linewidth}|}
\hline
\textbf{Research Dimensions}           & \textbf{Key Concerns}           & \textbf{Key Related Work} \\ \hline
Computation Optimization                 & Synchronization overhead, work efficiency                          &  CellJoin~\cite{CellJoin}, FPGAJoin~\cite{JoinFPGA,JoinFPGA2}, Handshake join~\cite{handshake,handshake_latency}, PanJoin~\cite{PanJoin},  HELLS-join~\cite{Karnagel2013StreamJP,Karnagel:2013:HHS:2485278.2485280}, Aggregation on GPU~\cite{SABER}, Aggregation on FPGA~\cite{agg_fpga,agg_fpga2}, Hammer Slide~\cite{Hammer}, StreamBox~\cite{StreamBox}, Parallel Index Join~\cite{ParallelIndex}              \\ \hline
Stream I/O Optimization                  & Time and space efficiency, data locality, and memory footprint & Batching~\cite{profile},   Stream with HBM~\cite{stream_hbm,stream_hbm2}, TerseCades~\cite{TerseCades}, Stream over InfiniBand~\cite{Heron_Infiniband}, Stream on SSDs~\cite{SSD_Stream}, and  NVM-aware Storage~\cite{NVM}             \\ \hline
Query Deployment                   & Operator interference, elastic scaling, and power constraint                          & Orchestrating~\cite{Orchestrating,Orchestrating2}, StreamIt~\cite{Carpenter:2009:MSP:1629395.1629406}, SIMD~\cite{Hormati:2010:MMS:1736020.1736053}, BitStream~\cite{bitStream}, Streams on Wires~\cite{mueller2009streams}, HRC~\cite{HRC}, RCM~\cite{RCM}, CMGG~\cite{CMGG}, GStream~\cite{GStream}, SABER~\cite{SABER}, BriskStream~\cite{briskstream}              \\ \hline
\end{tabular}%
}
\end{table*}
\section{Survey Outline}
\label{sec:overview}
The hardware architecture is evolving fast and provides a much higher processing capability than that traditional DSPSs were originally designed for.
For example, recent \emph{scale-up} servers can accommodate hundreds of CPU cores and terabytes of memory~\cite{sgi}, providing abundant computing resources. 
Emerging network technologies such as Remote Direct Memory Access (RDMA) and 10Gb Ethernet significantly improve system ingress rate, making I/O no longer a bottleneck in many practical scenarios~\cite{StreamBox,ingestion17}. 
However, prior studies~\cite{profile, Analyzing} have shown that existing data stream processing systems (DSPSs) severely underutilize hardware resources due to the unawareness of the underlying complex hardware architectures.

As summarized in Table~\ref{tbl:summary}, 
we are witnessing a revolution in the design of DSPSs that exploit emerging hardware capability, particularly along with the following three dimensions: 

\emph{1) Computation Optimization:}
Contrary to conventional DBMSs, 
there are two key features in DSPSs that are fundamental to many stream applications and computationally expensive: 
\emph{Windowing operation}~\cite{Agg_survey} (e.g., windowing stream join) and \emph{Out-of-order handling}~\cite{K-slack}. 
The former deals with infinite stream, and the latter handles stream imperfection.
The support for those expensive operations is becoming one of the major requirements for modern DSPSs and is treated as one of the key dimensions in differentiating modern DSPSs. 
Prior approaches use multicores~\cite{Hammer,StreamBox}, heterogeneous architectures (e.g., GPUs and Cell processors)~\cite{CellJoin,SABER}, and Field Programmable Gate Arrays (FPGAs)~\cite{handshake, handshake_latency,JoinFPGA,JoinFPGA2,agg_fpga,agg_fpga2} for accelerating those operations.

\emph{2) Stream I/O Optimization:} 
Cross-operator communication~\cite{profile} is often a major source of overhead in stream processing.
Recent work has revealed that the overhead due to cross-operator communication is significant, even without the TCP/IP network stack~\cite{profile, Analyzing}. 
Subsequently, research has been conducted on improving the efficiency of data grouping (i.e., output stream shuffling among operators) using High Bandwidth Memory (HBM)~\cite{stream_hbm}, compressing data in transmission with hardware accelerators and applying computation directly over compressed data~\cite{TerseCades}, and leveraging InfiniBand for faster data flow~\cite{Heron_Infiniband}.
Having said that, there are also cases where the application needs to temporarily store data for future usage~\cite{state_survey} (i.e., state management~\cite{flinkstate}). 
Examples include stream processing with large window operation (i.e., workload footprint larger than memory capacity) and stateful stream processing with high availability (i.e., application states are kept persistently). 
To relieve the disk I/O overhead, recent work has investigated how to achieve more efficient state management, leveraging   SSD~\cite{SSD_Stream} and non-volatile memory (NVM)~\cite{NVM}.

\emph{3) Query Deployment:} 
At an even higher point of view, researchers have studied launching a whole stream application (i.e., a query) into various hardware architectures. 
Similar to traditional database systems, the goal of query deployment in DSPS is to minimize operator interference/ cross-operator communication, balance hardware resource utilization, and so on. 
The major difference compared to traditional database systems lies in their different problem assumptions, and hence in their system architectures (e.g., infinite input stream~\cite{Viglas:2002:RQO:564691.564697}, processing latency constraints~\cite{Heinze2014}, and unique cost function of streaming operators~\cite{motto,kolchinsky2018join}).
To take advantage of modern hardware, 
prior works have exploited various hardware characteristics such as cache-conscious strategies~\cite{bitStream},  FPGA~\cite{mueller2009streams}, and GPUs~\cite{HRC,RCM,CMGG,GStream}.
Recent works have also looked into supporting hybrid architectures~\cite{SABER} and NUMA~\cite{briskstream}.

\compact

\begin{figure*}[h]
	\centering
     \includegraphics[width=\textwidth]{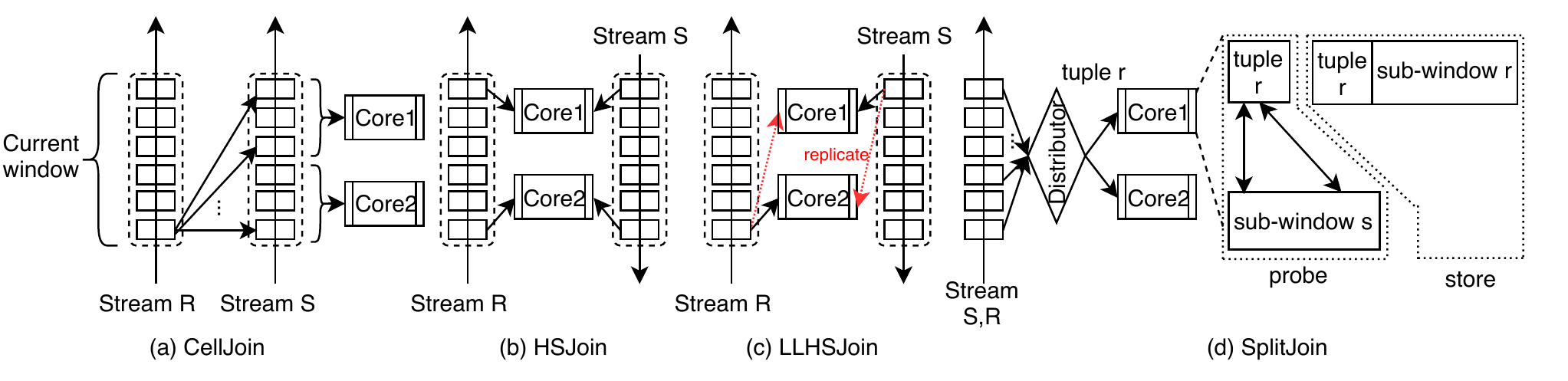}    
    \caption{HW-conscious stream join algorithms (using two cores as an example).}                   
    \label{fig:join}    
\end{figure*}

\section{Computation Optimization}
\label{sec:operator}
In this section, we review the literature on accelerating computationally expensive streaming operations using modern hardware.

\subsection{Windowing Operation}
In stream applications, the processing is mostly performed in the form of long-running queries known as continuous queries~\cite{cql}. 
To handle potentially infinite data streams, 
continuous queries are typically limited to a window that limits the number of tuples to process at any point in time. 
The window can be defined based on the number of tuples (i.e., count based), function of time (i.e., time based) or sessions~\cite{traub2019efficient}.
Window stream joins and window aggregation are two common expensive windowing operations in data stream processing. 

\subsubsection{Window Join}
A common operation used in many stream analytical workloads is to \emph{join} multiple data streams.
Different from traditional join in relational databases~\cite{Join_he}, which processes a large batch of data at once, 
stream join has to produce results on the fly~\cite{Kang2003EvaluatingWJ,Srivastava:2004:MEW:1316689.1316719,1214967,Golab:2003:PSW:1315451.1315495}. 
By definition, the stream join operator performs over infinite streams. 
In practice, streams are cut into finite slices/windows~\cite{SurveyJoin}. 
In a two-way stream join,
tuples from the left stream (\texttt{R})
are joined with tuples in the right stream (\texttt{S}) when the specified
key attribute matches, and the timestamp of tuples from both streams falls within the same window.

\textbf{Algorithm Overview.}
Kang et al.~\cite{Kang2003EvaluatingWJ} described the first streaming join implementations.
For each tuple $r$ of stream \texttt{R}, 
1) Scan the window associated with stream \texttt{S} and look for {matching} tuples;
2) {Invalidate} the old tuples in both windows;
3) {Insert} $r$ into the window of \texttt{R}.

\textbf{HW-Conscious Optimizations.}
The costly nature of stream join and the stringent response time requirements 
of stream applications have created significant interest in accelerating stream joining. 
Multicore processors that provide high processing capacity are ideal for executing costly windowed stream operators. 
However, fully exploiting the potential of a multicore processor is often challenging due to the complex processor microarchitecture, 
deep cache memory subsystem,
and the unconventional programming model in general.
Figure~\ref{fig:join} illustrates the four representative studies on accelerating window stream joins described as follows.

\emph{CellJoin:}
An earlier work from Gedik et al.~\cite{CellJoin}, called CellJoin, 
attempt to parallelize stream join on Cell processor, a heterogeneous multicore architecture.
CellJoin generally follows Kang's~\cite{Kang2003EvaluatingWJ} three-step algorithm. 
To utilize multicores, it re-partitions \texttt{S}, and each resulting partition is assigned to an individual core. 
In this way, the matching step can be performed in parallel on multiple cores.
A similar idea has been adopted in the work by Karnagel et al.~\cite{Karnagel2013StreamJP} to utilize the massively parallel computing power of GPU.

\emph{Handshake-Join (HSJoin):}
CellJoin essentially turns the join process into a scheduling and placement process. 
Subsequently, it is assumed that the window partition and fetch must be performed in global memory. 
The repartition and distribution mechanism essentially reveals that CellJoin generally follows the BSP model (see Section~\ref{subsec:archi}).
This is later shown to be ineffective when the number of cores is large~\cite{handshake}, and a new stream join technique called handshake join (i.e., HSJoin) was proposed. 
In contrast to CellJoin, HSJoin adopts the CO model. 
Specifically, both input streams notionally flow through the stream processing engine in opposite directions. 
As illustrated in Figure~\ref{fig:join} (b), the two sliding windows are laid out side by side, and predicate evaluations are continuously performed along with the windows whenever two tuples encounter each other. 

\emph{Low-Latency Handshake-Join (i.e., LLHSJoin):}
Despite its excellent scalability, the downside of HSJoin is that tuples may have to be queued for long periods of time before the match, resulting in high processing latency. 
In response, Roy et al.~\cite{handshake_latency} propose a \emph{low-latency} handshake-join (i.e., LLHSJoin) algorithm. 
The key idea is that, instead of sequentially forwarding each tuple through a pipeline of processing units, tuples are replicated and forwarded to all involved processing units (see the red dotted lines in Figure~\ref{fig:join} (c)) before the join computation is carried out by one processing unit (called a home node). 

\emph{SplitJoin:}
The state-of-the-art windowing join implementation called SplitJoin~\cite{SplitJoin} parallelizes the join process via the CO model. 
Rather than forwarding tuples bidirectionally, as in HSJoin or LLHSJoin, SplitJoin broadcasts each newly arrived tuple $t$ (from either \texttt{S} or \texttt{R}) to all processing units. 
In order to make sure that  each tuple is processed only once, $t$ is retained in exactly one processing unit chosen in a round-robin manner. 
Although SplitJoin~\cite{SplitJoin} and HSJoin~\cite{handshake} can achieve the same concurrency theoretically without any central coordination, the former achieves a much lower latency due to the linear chaining delay of the HSJoin. 
While LLHSJoin~\cite{handshake_latency} reduces the processing latency of HSJoin~\cite{handshake} by using a fast forwarding mechanism, it complicates the processing logic and reintroduces central coordination to the processing~\cite{SplitJoin}. 



\subsubsection{Window Aggregation}
Another computationally heavy windowing operation is window aggregation, which summarizes the most recent information in a data stream. 
There are four workload characteristics~\cite{traub2019efficient} of stream aggregation 
including 1) window type, which refers to the logic based on which system derives finite windows from a continuous stream, such as tumbling, sliding, and session;
2) windowing measures, which refers to ways to measure windows, such as time-based, count-based, and any other arbitrary advancing measures;
3) aggregate functions with different algebraic properties~\cite{Agg_survey2} such as \emph{invertible}, \emph{associative}, \emph{commutative}, and \emph{order-preserving};
and 4) stream (dis)order, which shall be discussed in Section~\ref{subsec:determ}.

\textbf{Algorithm Overview.}
The trivial implementation is to perform the aggregation calculation from scratch for every arrived data. 
The complexity is hence $O(n)$, where $n$ is the window size.
Intuitively, efficiently leveraging previous calculation results for future calculation is the key to reducing computation complexity, 
which is often called incremental aggregation. 
However, the effectiveness of incremental aggregation depends heavily on the aforementioned workload characteristics such as the property of the aggregation function.
For example, 
when the aggregation function is invertible (e.g., sum), we can simply update (i.e., increase) the aggregation results when a new tuple is inserted into the window and evict with the time complexity of $O(1)$. 
For faster answering median like function, which has to keep all the relevant inputs, instead of performing a sort on the window for each newly inserted tuple, 
one can maintain an \emph{order statistics tree} as auxiliary data structure~\cite{OST}, which has $O(log n)$ worst-case complexity of its insertion, deletion, and rank function. 
Similarly, the reactive aggregator (RA)~\cite{RA} with $O(log n)$ average complexity only works for aggregation function with the associative property.
 Those algorithms also differ from each other at their capability of handling different window types, windowing measures, and stream (dis)order~\cite{traub2019efficient}. 
 Traub et al.~\cite{traub2019efficient} recently proposed a generalization of the stream slicing technique to handle different workload characteristics for window aggregation. It may be an interesting future work to study how the proposed technique can be applied to better utilize modern hardware architectures (e.g., GPUs). 

\textbf{HW-Conscious Optimizations.}
There are a number of works on accelerating windowing aggregation in a hardware-friendly manner.
An early work by Mueller et al.~\cite{Mueller:2009:DPF:1687627.1687730} described implementation for a sliding windowed median operator on FPGAs. 
This is an operator commonly used to, for instance, eliminate noise in sensor readings and in data analysis tasks~\cite{1162749}.
The algorithm skeleton adopted by the work is rather conventional: it first sorts elements within the sliding window and then computes the median. 
Compared to the $O(log n)$ complexity of using an order statistics tree as an auxiliary data structure~\cite{OST}, Mueller's method has a theoretically much higher complexity due to the sorting step ($O(nlogn)$).
Nevertheless, their key contribution is on how the sorting and computing steps can be efficiently performed on FPGAs. 
Mueller's implementation~\cite{Mueller:2009:DPF:1687627.1687730} focuses on efficiently processing one sliding window without discussing how to handle subsequent sliding windows. 
Mueller et al. hence proposed conducting multiple computations for each sliding window by instantiating multiple aggregation modules concurrently~\cite{mueller2009streams}.

Recomputing from scratch for each sliding window is costly, even if conducted in parallel~\cite{mueller2009streams}. 
Hence, a technique called pane~\cite{pane} was proposed and later verified on FPGAs~\cite{agg_fpga} to address this issue.
The key idea is to divide overlapping windows into disjoint panes, compute sub-aggregates over each pane, and ``roll up" the partial-aggregates to compute final results.
Pane was later improved~\cite{pairs} and covers more cases (e.g., to support non-periodic windows~\cite{Cutty}). 
However, the latest efforts are mostly theoretical, and little work has been done to validate the effectiveness of these techniques on modern hardware, e.g., FPGA and GPUs.

Saber~\cite{SABER} is a relational stream processing system targeting
heterogeneous machines equipped with CPUs and GPUs.
To achieve high throughput, Saber also adopts incremental aggregation computations utilizing
the commutative and associative property of some aggregation functions such as count, sum, and average. 
Theodorakis et al.~\cite{Hammer} recently studied the trade-off between workload complexity and CPU efficient streaming window aggregation. 
To this end, they proposed an implementation that is both workload- and CPU-efficient. 
Gong et al.~\cite{ShuntFlow} proposed an efficient and scalable accelerator based on FPGAs, called ShuntFlow, to support arbitrary window sizes for both reduce- and index-like sliding window aggregations.
The key idea is to partition aggregation with extremely large window sizes into sub-aggregations with smaller window sizes that can enable more efficient use of FPGAs.


\subsection{Out-of-Order Handling} 
\label{subsec:determ}
In a real production environment, out-of-order\footnote{Other issues such as delay and missing can be seen as special cases of out-of-order.} input data are not uncommon. 
A stream operator is considered to be \emph{order-sensitive} if it requires input events to be processed in a certain predefined order (e.g., chronological order).
Handling out-of-order input data in an {order-sensitive} operator often turns out to be a performance bottleneck, as there is a fundamental conflict between data parallelism and order-sensitive processing -- the former seeks to improve the throughput of an operator by letting more than one thread operate on different events concurrently, possibly out-of-order. 

\textbf{Algorithm Overview.}
Currently, there are three general techniques to be applied together with the order-sensitive operator to handle out-of-order data streams.
The first utilizes a buffer-based data structure~\cite{K-slack} that buffers incoming tuples for a while before processing. 
The key idea is to keep the data as long as possible (within the latency/buffer size constraint) to avoid out-of-order inputs. 
The second technique relies on punctuation~\cite{OOP}, which is a special tuple in the event stream indicating the end of a substream. 
Punctuations guarantee that tuples are processed in monotonically increasing time sequence across punctuations, but not within the same punctuation.
The third approach is to use speculative techniques~\cite{Revision}. 
The main idea is to process tuples without any delay, and recompute the results in the case of order violation. 
There are also techniques specifically designed for handling out-of-order in a certain type of operator such as window aggregation~\cite{traub2019efficient}.

\begin{figure}
    \centering
     \includegraphics[width=.5\textwidth]{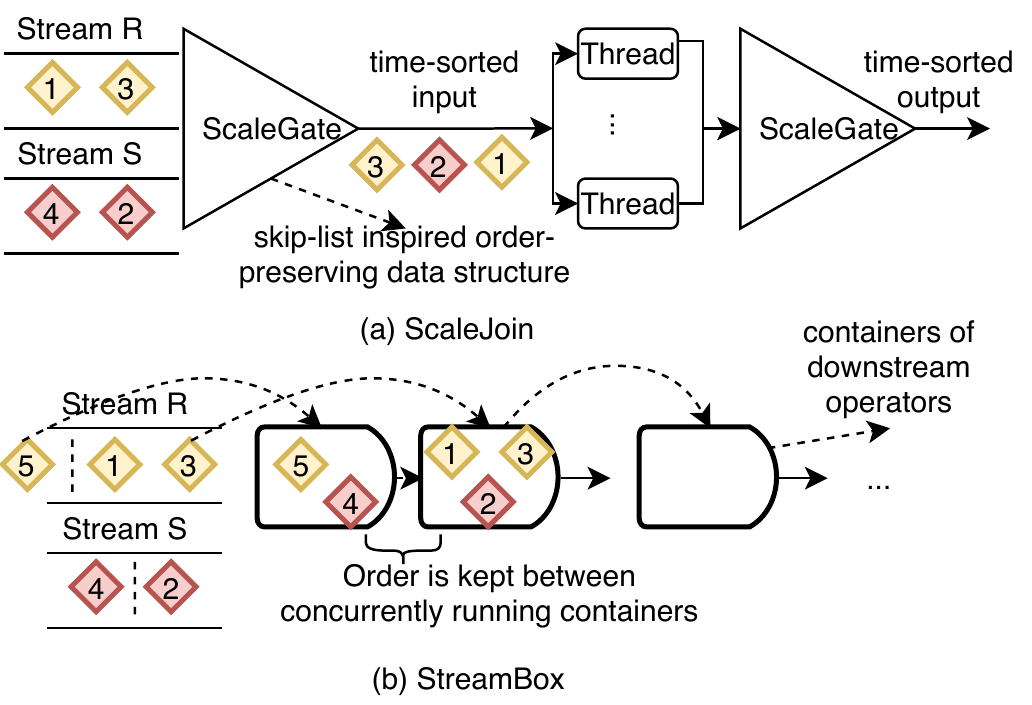}    
    \caption{Multicore-friendly out-of-order handling.}                   
    \label{fig:determ}    
\end{figure}

\textbf{HW-Conscious Optimizations.}
Gulisano et al.~\cite{Scalejoin} are among the first to handle out-of-order for high-performance stream join on multi-core CPUs.
The proposed algorithm, called \emph{scalejoin} is illustrated in Figure~\ref{fig:determ} (a). 
It first merges all incoming tuples into one stream (through a data structure called \emph{scalegate}) and then distributes them to processing threads (PTs) to perform join. 
The output also needs to be merged and sorted before exiting the system. 
The use of the \emph{scalegate} makes this work fall into the category of buffer-based approach and have inherent limitation of higher processing latency.
Scalejoin has been implemented in FPGA~\cite{JoinFPGA} and further improved in another recent work~\cite{JoinFPGA2}.
They both found that the proposed system outperforms the corresponding fully optimized parallel software-based solution running on a high-end 48-core multiprocessor platform.

StreamBox~\cite{StreamBox} handles out-of-order event processing by the punctuation-based technique on multicore processors. 
Figure~\ref{fig:determ} (b) illustrates the basic idea of taking the stream join operator as an example.  
Relying on a novel data structure called \emph{cascading container} to track dependencies between epochs (a group of tuples delineated by punctuation), 
StreamBox is able to maintain the processing order among multiple concurrently executing containers that exploit the parallelism of modern multicore hardware. 

Kuralenok et al.~\cite{kuralenok2018optimistic} attempt to balance the conflict between order-sensitive and multicore parallelism with an optimistic approach falling in the third approach. 
The basic idea is to conduct the joining process without any regulations, but apologize (i.e., sending amending signals) when the processing order is violated.
They show that the performance of the proposed approach depends on how often reorderings are observed during run-time. 
In the case where the input order is naturally preserved, there is almost no overhead. 
However, it leads to extra network traffic and computations when reorderings are frequent. 
To apply such an approach to practical use cases, it is hence necessary to predict the probability of reordering, which could be an  interesting future work.

\subsection{Remarks}
From the above discussion, it is clear that the key to accelerating windowing operators are mainly two folds. 
On the one hand, we should minimize the operation complexity. 
There are two common approaches: 
1) incremental computation algorithms~\cite{SABER}, which maximize  reusing intermediate results, and 2) rely on efficient auxiliary data structures (e.g., indexing the contents of sliding window~\cite{Ya-xin2006}) for reducing data (and/or instruction) accesses, especially cache misses.
On the other hand, we should maximize the system concurrency~\cite{Hammer}. 
This requires us to distribute workloads among multiple cores 
and minimize synchronization overhead among them~\cite{StreamBox}.
Unfortunately, these optimization techniques are often at odds with each other. 
For example, incremental computation algorithm is complexity efficient but difficult to parallelize due to inherent control dependencies in the CPU instruction~\cite{Hammer}. 
Another example is that maintaining index structures for partial computing results may help to reduce data accesses, but it also brings maintenance overhead~\cite{Lin:2015:SDS:2723372.2746485}.
More investigation is required to better balance these conflicting aspects.

\compact
\section{Stream I/O Optimization}
\label{sec:flow}
In this section, we review the literature on improving the stream I/O efficiency using modern hardware.

\subsection{Cross-operator Communication}
Modern
DSPSs~\cite{flink, Storm} are able to achieve very low processing latency in the order of milliseconds. 
However, excessive communication among operators~\cite{profile} is still a key obstacle in further improving the performance of the DSPSs. 

Kamburugamuve et al.~\cite{Heron_Infiniband} recently presented their findings on integrating Apache Heron~\cite{heron} 
with InfiniBand and Intel OmniPath. 
The results show that both can be utilized to improve the performance of distributed streaming applications. 
Nevertheless, many optimization opportunities remain to be explored. 
For example, prior work~\cite{Heron_Infiniband} has evaluated Heron on InfiniBand with channel semantics but not remote direct memory access (RDMA) semantics~\cite{infiniband}. The latter has shown to be very effective in other related works~\cite{Shi:2016:FCR:3026877.3026902, DistributedTxn}.

Data compression is a widely used approach for reducing communication overhead. 
Pekhimenko et al.~\cite{TerseCades} recently examined the potential of using data compression in stream processing. 
Interestingly, they found that data compression does not necessarily lead to a performance gain. 
Instead, improvement can only be achieved through a combination of hardware accelerator (i.e., GPUs in their proposal) and new execution techniques (i.e., compute directly over compressed data). 

As mentioned before, word count requires the same word to be transmitted to the same \texttt{Counter} operator (see Section~\ref{subsec:api}).
Subsequently, all DSPSs need to implement data grouping operations
regardless of their processing model (i.e., the continuous operator model or bulk synchronous model). 
Data grouping involves excessive memory accesses that rely on hash-based data structures~\cite{profile, SparkStreaming}.
Zeuch et al.~\cite{Analyzing} analyzed the design space of DSPSs optimized for modern multicore processors. 
In particular, they show that a queue-less execution engine based on query compilation, 
which replaces communication between operators with function calls, is highly suitable for modern hardware. 
Since data grouping can not be completely eliminated, they proposed a mechanism called ``Upfront Partitioning with Late Merging'', for efficient data grouping.
Miao et al.~\cite{stream_hbm} have exploited the possibility of accelerating data grouping using emerging 3D stacked memories such as high-bandwidth memory (HBM).
By designing the system in a way that addresses the limited capacity of HBM and HBM’s need for sequential-access and high parallelism, 
the resulting system can achieve several times of performance improvement over the baseline.

\subsection{State Management}
Emerging stream applications often require the
underlying DSPS to maintain large application states so as to
support complex real-time analytics~\cite{state_survey,flinkstate}. 
Representative example states required during stream processing include graph data structures~\cite{Wukong+S} and transaction records~\cite{S-Store}. 

The storage subsystem has undergone tremendous innovation in order to keep up with the ever-increasing performance demand. 
Wukong+S~\cite{Wukong+S} is a recently proposed distributed
streaming engine that provides real-time consistent query over streaming datasets.
It is built based on Wukong~\cite{Shi:2016:FCR:3026877.3026902}, which leverages RDMA to optimize throughput and latency. Wukong+S also follows its pace to support stream processing while maintaining low latency and high throughput.
Non-Volatile Memory (NVM) has emerged as a promising hardware and brings many new opportunities and challenges.
Fernando et al.~\cite{NVM} has recently explored efficient approaches to support analytical workloads on NVM, where an NVM-aware storage layout for tables is presented based on a multidimensional clustering approach and a block-like structure to utilize the entire memory stack.
As argued by the author, the storage structure designed on NVM may serve as the foundation for supporting features like transactional stream processing systems~\cite{TSM_NVM} in the future.
Non-Volatile Memory Express (NVMe) -based solid-state devices (SSDs) are expected to deliver unprecedented performance in terms of latency and peak bandwidth. 
For example, the recently announced PCIe 4.0 based NVMe SSDs~\cite{ssds} are already capable of achieving a peak bandwidth of 4GB/s.
Lee et al.~\cite{SSD_Stream} have recently investigated the performance limitations
of current DSPSs on managing application states on SSDs and have shown that query-aware optimization can significantly improve the performance of stateful stream processing on SSDs. 

\subsection{Remarks}
Hardware-conscious stream I/O optimization is still in its early days. 
Most prior work attempts at mitigating the problem through a purely software approach, such as I/O-aware query deployment~\cite{T-storm}.
The emerging hardware such as Non-Volatile Memory (NVM) and InfiniBand with RDMA open up new opportunities for further improving stream I/O performance~\cite{Wukong+S}. 
Meanwhile, the usage of emerging hardware accelerators such as GPUs further brings new opportunities to trade-off computation and communication overhead~\cite{TerseCades}.
However, a model-guided approach to balance the trade-off is still generally missing in existing work.
We hence expect more work to be done in this direction in the near future.

\compact

\section{Query Deployment}
\label{sec:query}
We now review prior works from a higher level of abstraction,
the query/application dimension. 
We summarize them based on their deployment targets: multicore CPUs, GPUs, and FPGAs.

\subsection{Multicore Stream Processing}
\textbf{Language and Compiler.}
Multicore architectures have become ubiquitous.
However, programming models and compiler techniques for employing multicore features are still lagging behind hardware improvements.
Kudlur et al.~\cite{Orchestrating} were among the first to develop a compiler technique to map stream application to a multicore processor. 
By taking the Cell processor as an example, they study how to compile and run a stream application expressed in their proposed language, called \texttt{StreamIt}. 
The compiler works in two steps: 1) operator fission optimization (i.e., split one operator into multiple ones) and 2) assignment optimization (i.e., assign each operator to a core). 
The two-step mapping is formulated as an integer linear programming (ILP) problem and requires a commercial ILP solver. 
Noting its NP-Hardness, Farhad et al.~\cite{Orchestrating2} later presented an approximation algorithm to solve the mapping problem.
Note that the mapping problem from Kudlur et al.~\cite{Orchestrating} considers only CPU loads and ignores communications bandwidth.
In response, Carpenter et al.~\cite{Carpenter:2009:MSP:1629395.1629406} developed an algorithm that maps a streaming program onto a heterogeneous target, further taking communication into consideration.
To utilize a SIMD-enabled multicore system, 
Hormati et al.~\cite{Hormati:2010:MMS:1736020.1736053} proposed vectorizing  stream applications.
Relying on high-level information, such as the relationship between operators, they were able to achieve better performance than general vectorization techniques. 
Agrawal et al.~\cite{bitStream} proposed a cache conscious scheduling algorithm for mapping stream application on multicore processors.
In particular, they developed the theoretical lower bounds on cache misses when scheduling a streaming pipeline on multiple processors, and the upper bound of the proposed cache-based partitioning algorithm called $seg\_cache$.
They also experimentally found that scheduling solely based on the cache effects can often be more effective than the conventional load-balancing (based on computation cost) approaches. 

\textbf{Multicore-aware DSPSs.}
Recently, there has been a fast growing amount of interest in building multicore-friendly DSPSs.
Instead of statically compiling a program as done in \texttt{StreamIt}~\cite{Orchestrating,Orchestrating2,Carpenter:2009:MSP:1629395.1629406}, these
DSPSs provide better elasticity for application execution. 
They also allow the usage of general-purpose programming languages (e.g., Java, Scala) to express stream applications.
Tang et al.~\cite{tang2013autopipelining} studied the data flow graph to explore the potential parallelism in a DSPS and proposed an \emph{auto-pipelining} solution that can utilize multicore processors to improve the throughput of stream processing applications.
For economic reasons, power efficiency has become more and more important in recent years,
especially in the HPC domains.
Kanoun et al.~\cite{kanoun2014low} proposed a multicore scheme for stream processing that takes power constraints  into consideration.
Trill~\cite{trill} is a single-node query processor for temporal or streaming data. 
Contrary to most distributed DSPSs (e.g., Storm, Flink) adopting the continuous operator model, Trill runs the whole query only on the thread that feeds data to it. Such an approach has shown to be especially effective~\cite{Analyzing} when applications contain no synchronization barriers.

\subsection{GPU-Enabled Stream Processing}
GPUs are the most popular heterogeneous processors due to their high computing  capacity.
However, due to their unconventional execution model, 
special designs are required to efficiently adapt stream processing to GPUs.

\textbf{Single-GPU.}
Verner et al.~\cite{HRC} presented a general algorithm for processing data streams with real-time stream scheduling constraints on GPUs.
This algorithm assigns data streams to CPUs and GPUs based on their incoming rates.
It tries to provide an assignment that can satisfy different requirements from various data streams. 
Zhang et al.~\cite{ZHANG201544} developed a holistic approach to building  DSPSs using GPUs.
They design a latency-driven GPU-based framework, which mainly focuses on real-time stream processing.
%
%
Due to the limited memory capacity of GPUs, 
the window size of the stream operator plays an important role in system performance.
Pinnecke et al.~\cite{pinnecke2015toward} studied the influence of window size
and proposed a partitioning method for splitting large windows into different batches, considering both time and space efficiency.
SABER~\cite{SABER} is a window-based hybrid stream processing framework aiming to utilize  CPUs and GPUs concurrently.

\textbf{Multi-GPU.}
Multi-GPU systems provide tremendous computation capacity, but also pose challenges like how to partition or schedule workloads among GPUs.
Verner et al.~\cite{RCM} extend their method~\cite{HRC} to a single node with multiple GPUs.
A scheduler controls stream placement and guarantees that the requirements among different streams can be met.
GStream~\cite{GStream} is the first data streaming framework for GPU clusters. 
GStream supports stream processing applications in the form of a C++ library;
it uses MPI to implement the data communication between different nodes
and uses CUDA to conduct stream operations on GPUs.
Alghabi et al.~\cite{alghabi2015scalable} first introduced the concept of stateful stream data processing on a node with multiple GPUs.
Nguyen et al.~\cite{CMGG} considered the scalability with the number of GPUs on a single node,
and developed a GPU performance model for stream workload partitioning in multi-GPU platforms with high scalability.
Chen et al.~\cite{chen2015g} proposed G-Storm,
which enables Storm~\cite{Storm} to utilize GPUs 
and can be applied to various applications that Storm has already supported.

\subsection{FPGA-Enabled Stream Processing}
FPGAs are programmable integrated circuits whose hardware interconnections can be configured by users.
Due to their low latency, high energy efficiency, and low hardware engineering cost,
FPGAs have been explored in various application scenarios, including stream processing.

Hagiescu et al.~\cite{hagiescu2009computing} first elaborated challenges to implementing stream processing on FPGAs and proposed algorithms 
that optimize processing throughput and latency for FPGAs.
Mueller et al.~\cite{mueller2009streams} provided \emph{Glacier},
which is an FPGA-based query engine that can process queries on streaming data from networks.
The operations in \emph{Glacier} include selection, aggregation, grouping, and windows.
Experiments show that using FPGAs helps achieve much better performance than using conventional CPUs.
A common limitation of an FPGA-based system is its expensive synthesis process, which takes a significant time to compile the application into hardware designs for FPGAs. 
This makes FPGA-based systems inflexible in adapting to query changes.
In response, Najafi et al.~\cite{najafi2013flexible} demonstrated Flexible Query Processor (FQP), an
online reconfigurable event stream query processor that can accept new queries without disrupting other queries in execution.

\subsection{Remarks}
Existing systems usually involve heterogeneous processors along with CPUs. 
Such heterogeneity opens up both new opportunities and poses challenges for scaling stream processing. 
From the above discussion, it is clear that both GPUs and FPGAs have been successfully applied for scaling up stream processing. 
FPGAs have low latency and are hardware configurable. 
Hence, they are suitable for special application scenarios, such as a streaming network. 

\compact

\section{System Design Requirements}
\label{sec:oppo}
In 2005, Stonebraker et al.~\cite{requirements} outlines  eight requirements of real-time data stream processing. 
Since then, tremendous improvements have  been made thanks to the great efforts from both industry and the research community. 
We now summarize how hardware-conscious optimization techniques mitigate the gap between DSPSs and requirements while highlighting the insufficiency.

Most DSPSs are designed with the principle of ``\emph{Keep the Data Moving}"~\cite{requirements}, 
and hence aim to process input data ``on-the-fly'' without storing them. 
As a result, message passing is often a key component in the  current DSPSs.
To mitigate the overhead, researchers have recently attempted to improve the cross-operator communication efficiency by taking advantage of the latest advancement in network infrastructure~\cite{Heron_Infiniband}, compression using hardware accelerator~\cite{TerseCades}, and efficient algorithms by exploiting new hardware characteristics~\cite{stream_hbm}. 
Going forward, we expect more work  to be done for hardware-conscious stream I/O optimization.

Handling out-of-order input streams is relevant to both the \emph{Handle Stream Imperfections} and 
\emph{Generate Predictable Outcomes}~\cite{requirements} requirements. 
In real-time stream systems where the input data are not stored,
the infrastructure must make provision for handling data
that arrive late or are  delayed, missing or out-of-sequence.
Correctness can be guaranteed in some applications only if time-ordered and 
deterministic
processing is maintained throughout the entire processing pipeline. 
Despite the significant efforts, existing DSPSs are still \emph{far from ideal} for  exploiting the potential of modern hardware. 
For example, as observed in a recent work~\cite{briskstream}, the same DSPS (i.e., StreamBox) delivers much lower throughput on modern multicore processors as a result of enabling ordering guarantees.

The state management in DSPSs is more related to the \emph{Integrate Stored and Streaming Data}~\cite{requirements} requirement.
For many stream processing applications, comparing the ``present'' with the ``past'' is a common task. 
Thus, the system must provide careful management of the stored states. 
However, we observe that only a few related studies attempt to improve state management efficiency levering modern hardware~\cite{SSD_Stream}. 
There are still many open questions to be resolved, such as new storage formats, indexing techniques  for emerging hardware architectures and applications~\cite{TSM_NVM,report}. 
New media applications such as live audio streaming services~\cite{8509389} also challenge existing systems in terms of new processing paradigms. 

The \emph{Partition and Scale Applications Automatically}~\cite{requirements} 
requires a DSPS to be able to elastically scale up and down in order to process input streams with varying characteristics.
However, based on our analysis, little work has considered scaling \emph{down} the processing efficiently (and easily scaling up later) in a hardware-aware manner. 
A potential direction is adopting a server-less computing paradigm~\cite{Baldini2017} with the help of novel memory techniques such as Non-Volatile Memory (NVM) into DSPSs. 
However, questions such as how to efficiently manage the partial computing state in GPUs or FPGAs still remain unclear.

The proliferation of high-rate data sources has accelerated the development of next-generation performance-critical DSPSs. 
For example, the new 5G network promises blazing speeds, massive throughput capability, and ultra-low latencies~\cite{5g_report}, thus bringing the higher potential for performance critical stream applications.
In response, high-throughput stream processing is essential to keeping up with data streams in order to satisfy the \emph{Process and Respond Instantaneously}~\cite{requirements} requirement.
However, achieving high-throughput stream processing is challenging, especially when expensive windowing operations are deployed. 
By better utilizing modern hardware, researchers and practitioners have achieved promising results. For example, SABER processes $79$ million tuples per second with eight  CPU cores for Yahoo Streaming Benchmark, outperforming other DSPSs several times~\cite{saber_blog}. 
Nevertheless, current results also show that there is still room for improvement on a single node, and this constitutes an opportunity for  designing the next-generation DSPSs~\cite{ZeuchCdMGGGBTM2019}.

Two requirements including \emph{Query using SQL on Streams} and \emph{Guarantee Data Safety and Availability}  are overlooked by most existing HW-conscious optimization techniques in DSPSs. 
In particular, how to design HW-aware SQL statements for DSPSs, and how best to guarantee data safety and system availability when adopting modern hardware, such as NVM for efficient local backup and high-speed network for remote backup, remain an open question.


%


\compact
\section{Conclusion}
\label{sec:conclude}
In this paper, we have discussed relevant literature from the field of hardware-conscious DSPSs, which aim to utilize modern hardware capabilities for accelerating stream processing. 
Those works have significantly improved DSPSs to better satisfy the design requirements raised by Stonebraker et al.~\cite{requirements}.
%
In the following, we list some additional advice on future research directions. 

\textbf{Scale-up and -out Stream Processing.}
As emphasized by Gibbons~\cite{Gibbons}, scaling both out and up is crucial to effectively improving  the system performance.
In situ analytics enable data processing at the point of data origin, thus reducing the data movements across   networks; 
Powerful hardware infrastructure provides an opportunity to improve processing performance within a single node. 
To this end, many recent works have exploited the potential of high-performance stream processing on a single node~\cite{SABER, StreamBox, Analyzing}.
However, the important question of how best to use powerful local nodes in the context of large distributed computation setting still remains unclear. 

\textbf{Stream Processing Processor.}
With the wide adoption of stream processing today, it may be a good time to revisit the design of a specific processor for DSPSs.  
GPUs~\cite{SABER} provide much higher bandwidth  than CPUs, but it comes with larger latency as tuples must be first accumulated in order to fully utilize thousands of cores on GPU;
FPGA~\cite{JoinFPGA} has its advantage in providing low latency, low power consumption computation but its throughput is still much lower compared to GPUs.
The requirement for an ideal processor for stream processing includes \emph{low latency}, \emph{low power consumption}, and \emph{high bandwidth}. 
On the other hand, components like complex control logic may be sacrificed as stream processing logic is usually predefined and fixed. Further, due to the nature of continuous query processing, it is ideal to keep the entire instruction set  close to processor~\cite{profile}.

\noindent\textbf{Acknowledgments.}
\begin{small}
The authors would like to thank the anonymous reviewer and the associate editor, Pınar Tözün, for their insightful comments on improving this manuscript.
This work is supported by a MoE Tier 1 grant (T1 251RES1824) and a MoE Tier 2 grant (MOE2017-T2-1-122) in Singapore.
Feng Zhang's work was  partially supported by the National Natural Science Foundation of China (Grant No. 61802412, 61732014).
\end{small}

\compact
\bibliographystyle{abbrv} 
{\small
\bibliography{mybib}
}

\end{document}